\documentclass[preprint,showpacs]{revtex4}
\usepackage{amsfonts}
\usepackage{amsmath}
\usepackage{amssymb}

\begin{document}

\title[GSL in gravity]{The generalized second law of thermodynamics in
generalized gravity theories}
\author{Shao-Feng Wu$^{1}$\footnote{%
Corresponding author. Email: sfwu@shu.edu.cn; Phone:
+86-021-66136202.}, Bin
Wang$^{2}$\footnote{%
Email: wangb@fudan.edu.cn}, Guo-Hong Yang$^{1}$\footnote{%
Email: ghyang@mail.shu.edu.cn}, and Peng-Ming Zhang$^{3,4}$\footnote{%
Email: zhpm@impcas.ac.cn}} \affiliation{$^{1}$Department of Physics,
Shanghai University, Shanghai, 200436, P. R. China}
\affiliation{$^{2}$Department of Physics, Fudan University, Shanghai
200433, P. R. China} \affiliation{$^{3}$Center of Theoretical
Nuclear Physics, National Laboratory of Heavy Ion Accelerator,
Lanzhou 730000, P. R. China} \affiliation{$^{4}$Institute of Modern
Physics, Lanzhou, 730000, P. R. China} 

\begin{abstract}
We investigate the generalized second law of thermodynamics (GSL) in
generalized theories of gravity. We examine the total entropy evolution with
time including the horizon entropy, the non-equilibrium entropy production,
and the entropy of all matter, field and energy components. We derive a
universal condition to protect the generalized second law and study its
validity in different gravity theories. In Einstein gravity, (even in the
phantom-dominated universe with a Schwarzschild black hole), Lovelock
gravity, and braneworld gravity, we show that the condition to keep the GSL
can always be satisfied. In $f(R)$ gravity and scalar-tensor gravity, the
condition to protect the GSL can also hold because the gravity is always
attractive and the effective Newton constant should be approximate constant
satisfying the experimental bounds.
\end{abstract}

\maketitle

\section{Introduction}

Motivated by the black hole physics, it was realized that there is a
profound connection between gravity and thermodynamics. In Einstein gravity,
the evidence of this connection was first discovered in \cite{Jacobson} by
deriving the Einstein equation from the proportionality of entropy and
horizon area together with the first law of thermodynamics in the Rindler
spacetime. For a general static spherically symmetric spacetime, Padmanabhan
pointed out that Einstein equations at the horizon give rise to the first
law of thermodynamics \cite{Padmanabhan}. Recently the study on the
connection between gravity and thermodynamics has been extended to
cosmological context. Frolov and Kofman \cite{Frolov} employed the approach
proposed by Jacobson \cite{Jacobson} to a quasi-de Sitter geometry of
inflationary universe, and calculated the energy flux of a background
slow-roll scalar through the quasi-de Sitter apparent horizon. By applying
the first law of thermodynamics to a cosmological horizon, Danielsson
obtained Friedmann equation in the expanding universe \cite{Danielsson}. In
the quintessence dominated accelerating universe, Bousso \cite{Bousso}
showed that the first law of thermodynamics holds at the apparent horizon.
The relation between gravity and thermodynamics has been further disclosed
in extended gravity theories, including Lovelock gravity \cite{Akbar1,Cao},
braneworld gravity \cite{Cao1,brane}, nonlinear gravity \cite%
{Eling,Akbar,Cao}, and scalar-tensor gravity \cite{Akbar,Cao} etc. In the
nonlinear gravity and scalar-tensor gravity, it was argued that the
non-equilibrium thermodynamics instead of the equilibrium thermodynamics
should be taken into account to build the relation to gravity \cite%
{Eling,Akbar,Cao}. In our previous work \cite{Wu1}, we have presented a
general procedure to build the connection between gravity and
thermodynamics. From the Friedmann equations, we have constructed the first
law of thermodynamics on the apparent horizon in generalized gravity
theories. We found that the non-equilibrium entropy production term arising
in non-linear gravity and scalar-tensor gravity is due to the existence of
other dynamic fields besides the ordinary matter dominating the cosmological
evolution.

It is of great interest to extend our discussion in \cite{Wu1} to study the
generalized second law (GSL) of thermodynamics in the generalized gravity
theories. There have been a lot of interest on investigating the GSL in
gravity \cite%
{Babichev,Setare,Pollock,Davies,Mohseni1,Izquierdo,Mohseni2,Wang,Zhou}, but
all of them concentrate on the Einstein gravity. The modified theory of
gravity was argued to be a possible candidate to explain the accelerated
expansion of our universe, thus it is interesting to examine the GSL in the
extended gravity theories. An attempt to study this problem was carried out
in \cite{Mohseni}, where it was found that some additional conditions are
needed for validity of GSL. Even for the Einstein gravity, it was found that
GSL breaks down in phantom-dominated universe in the presence of
Schwarzschild black hole \cite{Izquierdo}, at least in transition epoch \cite%
{Mohseni2}. In our paper we will adopt the formalism proposed in \cite{Wu1}.
We will derive the entropy of the horizon from the first law of
thermodynamics constructed in \cite{Wu1}. We will examine the total entropy
evolution with time including the horizon entropy, the non-equilibrium
entropy production, and the entropy of all matter, field and energy
components. We will derive a universal condition to protect the GSL in
generalized gravity theories and examine its validity in the Einstein
gravity (even in the presence of Schwarzschild black hole), Lovelock
gravity, braneworld gravity, nonlinear gravity and scalar-tensor gravity.

The organization of the paper is as follows: In section 2, we briefly review
the generalized first law of thermodynamics in extended theories of gravity.
In section 3, we derive the universal condition to protect the GSL and
examine its validity in some extended gravity theories. The last section is
devoted to summary.

\section{The first law of thermodynamics on the apparent horizon in FRW
cosmology}

In this section we briefly go over the general procedure to construct the
first law of thermodynamics on the apparent horizon in generalized gravity
theory \cite{Wu1}.

The homogenous and isotropic ($n+1$)-dimensional FRW universe is described by%
\begin{equation}
ds^{2}=h_{ab}dx^{a}dx^{b}+\tilde{r}^{2}d\Omega _{n-1}^{2},  \label{FRW}
\end{equation}%
where $h_{ab}=$diag$(-1,\frac{a^{2}}{1-ka^{2}})$, $d\Omega _{n-1}^{2}$ is
the $\left( n-1\right) $-dimensional sphere element, and $%
x^{0}=t,\;x^{1}=r,\;\tilde{r}=ar$ is the radius of the sphere and $a$ is the
scale factor. For simplicity, we consider the flat space $k=0$ in this
paper, however, our discussion can also be generalized to the non-flat
cases. It is known that the dynamical apparent horizon, the marginally
trapped surface with vanishing expansion, is defined as a sphere situated at
$r=r_{A}$ satisfying%
\begin{equation*}
h^{ab}\partial _{a}\tilde{r}\partial _{b}\tilde{r}=0.
\end{equation*}%
The sphere radius is%
\begin{equation}
\tilde{r}_{A}\equiv r_{A}a=\frac{1}{H}.  \label{Horizon}
\end{equation}%
The associated temperature on the apparent horizon is defined by surface
gravity $\kappa =\frac{1}{\sqrt{-h}}\partial _{a}\left( \sqrt{-h}%
h^{ab}\partial _{b}\tilde{r}\right) $%
\begin{equation}
T=\frac{\left\vert \kappa \right\vert }{2\pi }=\frac{1}{2\pi \tilde{r}_{A}}%
\left( 1-\epsilon \right)  \label{T}
\end{equation}%
where $\epsilon \equiv \frac{\partial _{t}\tilde{r}_{A}}{2H\tilde{r}_{A}}<1$%
. $\epsilon <1$ ensures that the temperature is positive. Using the
definition of the horizon (\ref{Horizon}), the positive temperature
condition can be written as $\epsilon =-\frac{\dot{H}}{2H^{2}}<1$, i.e.%
\begin{equation}
\dot{H}>-2H^{2},  \label{postive T}
\end{equation}%
which is useful in our later discussion. In our previous work \cite{Wu1}, in
order to study the mass-like function, we construct the first law on the
assumption $\epsilon \ll 1$ meaning that the apparent horizon radius is
approximately fixed thereby the temperature $T=\frac{1}{2\pi \tilde{r}_{A}}$
\cite{Cai,Cao}. Here we will drop this assumption.

In Einstein gravity, the entropy is proportional to the horizon area%
\begin{equation*}
S_{E}=\frac{A}{4G},
\end{equation*}%
where the horizon area $A=n\Omega _{n}\tilde{r}_{A}^{n-1}$. The
thermodynamical fluid $\delta Q$ can be written as%
\begin{equation*}
TdS_{E}=\frac{n(n-1)V\tilde{r}_{A}^{-3}d\tilde{r}_{A}}{8\pi G}-\frac{n(n-1)V%
\tilde{r}_{A}^{-3}d\tilde{r}_{A}}{8\pi G}\frac{\partial _{t}\tilde{r}_{A}}{2H%
\tilde{r}_{A}},
\end{equation*}%
where $V=\Omega _{n}\tilde{r}_{A}^{n}$ is the volume in the horizon. Using
the definition of the horizon (\ref{Horizon}) and the temperature (\ref{T}),
we can obtain%
\begin{equation}
TdS_{E}=\frac{-n(n-1)V}{16\pi G}\frac{dH^{2}}{dt}dt-\frac{n(n-1)V}{16\pi G}%
\frac{\dot{H}^{2}}{H}dt,  \label{dr1}
\end{equation}%
which is purely a geometric relation.

In all gravity theories, Friedmann equations can be expressed in the form as
that in the Einstein gravity%
\begin{equation}
H^{2}=\frac{16\pi G}{n(n-1)}\rho _{eff}  \label{H21}
\end{equation}%
\begin{equation}
\dot{H}=-\frac{8\pi G}{(n-1)}(\rho _{eff}+p_{eff}).  \label{H22}
\end{equation}%
Though we do not know the exact form of $\rho _{eff}$ (and $p_{eff}$), we
know that there must be ordinary matter density $\rho $ in $\rho _{eff}$ and
also other variables $\rho _{p}$. In some cases, $\rho _{p}$ (or their
combination) may describe other matter field $\rho _{f}$ or effective energy
component $\rho _{e}$ besides the ordinary matter $\rho $ in $\rho _{eff}$.
The first Friedmann equation can be expressed in the form%
\begin{equation*}
H^{2}=H^{2}(\rho ,\;\rho _{1},\cdots \rho _{p},\cdots ).
\end{equation*}%
Then the relation (\ref{dr1}) can be changed as%
\begin{equation}
TdS_{E}=\frac{-n(n-1)V}{16\pi G}dt(\frac{\partial H^{2}}{\partial \rho }\dot{%
\rho}+\frac{\partial H^{2}}{\partial \rho _{p}}\dot{\rho}_{p})-\frac{n(n-1)V%
}{16\pi G}\frac{\dot{H}^{2}}{H}dt.  \label{dr2}
\end{equation}%
To construct the first law of thermodynamics, we need to know the energy
flux $dE$ or entropy $S$. In the general gravity theory, they are not
specified. However, it is known that the energy flux of ordinary matter
includes $V\dot{\rho}dt$. Multiplying $\frac{16\pi G}{n(n-1)}\frac{1}{\frac{%
\partial H^{2}}{\partial \rho }}$ on both sides of (\ref{dr2}), we can
extract it clearly%
\begin{equation}
\frac{16\pi G}{n(n-1)}\frac{1}{\frac{\partial H^{2}}{\partial \rho }}%
TdS_{E}=-V\dot{\rho}dt-Vdt\frac{1}{\frac{\partial H^{2}}{\partial \rho }}%
\frac{\dot{H}^{2}}{H}-Vdt\frac{1}{\frac{\partial H^{2}}{\partial \rho }}%
\frac{\partial H^{2}}{\partial \rho _{p}}\dot{\rho}_{p}.  \label{dr4}
\end{equation}%
In the general case, we have the conservation%
\begin{equation}
\dot{\rho}_{eff}+nH(\rho _{eff}+p_{eff})=0.  \label{Con1}
\end{equation}%
We furthermore assume that ordinary matter (All matter and energy in this
paper are assumed as perfect fluid) has energy exchange with other energy
source, described by%
\begin{equation}
\dot{\rho}+nH(\rho +p)=q,  \label{Con2}
\end{equation}%
If the gravity theory has matter $\rho _{f}$ and energy content $\rho _{e}$,
one may also have similar semi-conserved laws%
\begin{align}
\dot{\rho}_{f}+nH(\rho _{f}+p_{f})& =q_{f},  \notag \\
\dot{\rho}_{e}+nH(\rho _{e}+p_{e})& =q_{e},  \notag \\
\dot{\rho}_{t}+nH(\rho _{t}+p_{t})& =q_{t}.  \label{qt}
\end{align}%
In the last equation the total density $\rho _{t}\equiv \rho +\rho _{f}+\rho
_{e}$, total pressure $p_{t}\equiv p+p_{f}+p_{e}$, and total energy exchange
$q_{t}\equiv q+q_{f}+q_{e}$ are introduced. However, it should be emphasized
that one can not impose total energy fluid $q_{t}=0$, because there may be
energy exchange with the horizon. Equations (\ref{Con1}) and (\ref{Con2})
will be used later to express the first law explicitly. Since we will
consider thermodynamical effect with the change of horizon volume, we
introduce the work density. Defining $T_{a}^{b}$ as the projection of
energy-momentum tensor $T_{\nu }^{\mu }$ of the perfect fluid in the FRW
universe in the normal direction of the ($n-1$)-sphere, we have the density $%
W\equiv -\frac{1}{2}T_{a}^{a},$ which may be viewed as the work done by the
change of the apparent horizon, as pointed out in \cite{Hayward}.

Consider the entropy change should be an exact form in the first law of
thermodynamics. If there is just ordinary matter $\rho $ in the space, $%
\frac{\partial H^{2}}{\partial \rho }$ can be rewritten as a function of $%
\tilde{r}_{A}$. Then a total differential can be obtained by the integration%
\begin{equation}
S=\int \frac{16\pi G}{n(n-1)}\frac{1}{\frac{\partial H^{2}}{\partial \rho }(%
\tilde{r}_{A})}d\left( S_{E}\right) ,  \label{S0}
\end{equation}%
and the relation (\ref{dr4}) can be written as%
\begin{equation}
TdS=\delta Q,  \label{first law1}
\end{equation}%
where the thermodynamical fluid%
\begin{equation}
\delta Q=-V\dot{\rho}dt-Vdt\frac{1}{\frac{\partial H^{2}}{\partial \rho }}%
\frac{\dot{H}^{2}}{H}.  \label{dE1}
\end{equation}%
Since the gravity is only determined by ordinary matter, we can impose
conservation $q=0$. Using the conservation equation (\ref{Con2}), one can
prove%
\begin{equation*}
-V\frac{1}{\frac{\partial H^{2}}{\partial \rho }}\frac{\dot{H}^{2}}{H}=-%
\frac{1}{2}\left( \rho +p\right) \dot{V}.
\end{equation*}%
Then the thermodynamical fluid can be rewritten as%
\begin{equation}
\delta Q=-dE+WdV,  \label{dq}
\end{equation}%
which includes just the energy flux of ordinary matter $E=\rho V$ and the
work done by the change of the apparent horizon. Thus the expression (\ref%
{S0}) should be understood as the entropy to assure the first law%
\begin{equation*}
TdS=-dE+WdV.
\end{equation*}%
\ \

If the modified gravity theory has other dynamic fields resulting that $%
\frac{\partial H^{2}}{\partial \rho }$ is a function of $\tilde{r}_{A}$ and $%
\rho _{p}$%
\begin{equation*}
\frac{\partial H^{2}}{\partial \rho }=\frac{\partial H^{2}}{\partial \rho }(%
\tilde{r}_{A},\rho _{p}),
\end{equation*}%
we can not integral the l.h.s in (\ref{dr4}) directly. However we can
express it as%
\begin{equation}
T\frac{16\pi G}{n(n-1)}\frac{1}{\frac{\partial H^{2}}{\partial \rho }}%
dS_{E}=Td\left( \frac{16\pi G}{n(n-1)}\frac{1}{\frac{\partial H^{2}}{%
\partial \rho }}S_{E}\right) -T\frac{16\pi G}{n(n-1)}S_{E}d\frac{1}{\frac{%
\partial H^{2}}{\partial \rho }}.  \label{first law20}
\end{equation}%
It can be rewritten as%
\begin{equation}
TdS=\delta Q,
\end{equation}%
where%
\begin{equation}
S\equiv \frac{16\pi G}{n(n-1)}\frac{1}{\frac{\partial H^{2}}{\partial \rho }}%
S_{E},  \label{S}
\end{equation}%
\begin{equation}
\delta Q\equiv -V\dot{\rho}dt-Vdt\frac{1}{\frac{\partial H^{2}}{\partial
\rho }}\frac{\dot{H}^{2}}{H}-Vdt\frac{1}{\frac{\partial H^{2}}{\partial \rho
}}\frac{\partial H^{2}}{\partial \rho _{p}}\dot{\rho}_{p}+T\frac{16\pi G}{%
n(n-1)}S_{E}d\frac{1}{\frac{\partial H^{2}}{\partial \rho }}.  \label{dE}
\end{equation}%
To construct the first law, we can write%
\begin{equation*}
\delta Q=-dE+W_{t}dV-Td_{p}S,
\end{equation*}%
where $E\equiv \rho _{t}V$ is the total intrinsic energy and $W_{t}$ the
total work. $d_{p}S$ is defined as%
\begin{equation}
d_{p}S\equiv -\frac{1}{T}\left( \delta Q+dE-W_{t}dV\right) \text{,}
\label{dis}
\end{equation}%
where the subindex \textquotedblleft p" denotes the term resulted from other
dynamical fields. In general, $d_{p}S$ is non-vanishing and can not be
written out exactly (We will show $d_{p}S$ clearly in concrete gravity
theories). If the first law of generalized gravity theory can be
constructed, the entropy term contains the form (\ref{S}) together with the
entropy production $d_{p}S$ (\ref{dis}) following the idea given in \cite%
{Eling}.\textbf{\ }Here we do not consider whether the entropy is developed
internally by the system as in \cite{Eling}. The first law is expressed as
\begin{equation}
TdS+Td_{p}S=-dE+W_{t}dV.  \label{first law2}
\end{equation}%
The exact forms of entropy production $d_{p}S$ and entropy $S$ depend on the
concrete gravity theory.

In \cite{Gong} it was argued that the entropy correction can be absorbed in
the mass-like function and the entropy generation in the nonequilibrium can
be reinterpreted. Whether one should interpret the thermodynamical equations
as representing systems in equilibrium or out of equilibrium seems unclear.
This is also under debate in recent works on $f(R)$ gravity \cite%
{Eling1,Elizalde}. In our case, there is work term, since the horizon is not
fixed as that in \cite{Gong} and the form of the mass-like function is not
known. The entropy expression, inner energy and the work term in the first
law Eq. (\ref{first law2}) are all determined. The extra term can not be
absorbed into any other terms and it has to be interpreted as the entropy
production as done in \cite{Eling}. Since the mass-like function here is not
available, we do not know how to reinterprete this entropy correction as
done in \cite{Gong}.

One might argue that there seems ambiguities in entropy expressions (\ref{S0}%
) (\ref{S}) since one might add a proper quantity to the expression of
entropy, which may vanish in the Einstein gravity, and this extra term could
be absorbed into the redefinition of the entropy production. This worry is
not necessary. As pointed out in \cite{Wu1}, the known black hole entropy in
different gravity theories will strictly restrict the form of the additional
quantities in the entropy expressions. The definition of entropy in (\ref{S0}%
) and (\ref{S}) can recover the exact expression of the known black hole
entropy \cite{Wu1} such as in Lovelock gravity \cite{Cai1}, nonlinear
gravity \cite{Wald} and scalar-tensor gravity \cite{Cai2}. Moreover,
following the work in \cite{Gong}, the general mass-like function which can
be reduced to the corresponding Misner-Sharp mass has been found. If we add
other quantities in entropy expressions (\ref{S0}) (\ref{S}), it seems very
difficult to obtain the corresponding mass-like function which can be
reduced to the known Misner-Sharp mass. This serves as another restriction
on adding additional terms to the entropy expressions.

\section{GSL of extended gravity theories}

Now we start to discuss the GSL in generalized gravity theories. Most
discussions on the GSL focus on the Einstein gravity. In generalized gravity
theories, from the first law of thermodynamics (\ref{first law2}) one can
see that an entropy production term appears, which characterizes the
non-equilibrium thermodynamical process about the horizon. In describing the
GSL one should include this non-equilibrium entropy production term.
Besides, in addition to the matter and fields, one should also include the
entropy for the effective energy in describing the GSL, since the change of
energy is relative to the change of entropy.

Now we are going to use the first law of thermodynamics (\ref{first law2})
to find the general condition need to hold the GSL in any gravity theories.
From (\ref{first law2}), we have the expression of the horizon entropy%
\begin{equation}
TdS_{h}=-Td_{p}S-dE+W_{t}dV.  \label{Sh2}
\end{equation}%
\ \ On the other hand, if there is only ordinary matter inside the
cosmological horizon of a comoving observer, the entropy of the ordinary
matter is relative to the energy $\rho $ and the pressure $p$ via the Gibbs
equation%
\begin{equation}
T_{\rho }dS_{\rho }=d(\rho V)+pdV=Vd\rho +(\rho +p)dV.  \label{Sroi}
\end{equation}%
As pointed out in \cite{Pavon}, the ordinary matter should be understood as
a phenomenological representation of a mixture of fields, each of which may
or may not be in a pure state, and therefore entitled to an entropy. $%
T_{\rho }$ refers to the temperature of matter inside the horizon. If there
are other matter field and energy component, one can similarly have the
Gibbs equation (including all matter, field, and energy contents)
\begin{equation}
T_{t}dS_{t}=d(\rho _{t}V)+p_{t}dV=Vd\rho _{t}+(\rho _{t}+p_{t})dV.
\label{St}
\end{equation}%
Here the temperature of total energy inside the horizon is denoted as $T_{t}$%
. It should be noted that the temperature $T_{t}$ is not equal to $T$ in
general because there may be energy flow $q_{t}$ (thereby thermodynamical
fluid) between the horizon and energy inside the horizon. Since the only
temperature scale we have at our disposal is the temperature of the apparent
horizon $T$, we assume%
\begin{equation*}
T_{t}=bT,
\end{equation*}%
where the temperature parameter satisfying $0<b<1$ assures the temperature
being positive and smaller than the horizon temperature. This ansatz is
similar to the disposal when the horizon is taken as the event horizon \cite%
{Pavon, Mohseni1}, where the only temperature scale is the de Sitter
temperature \cite{Davies}.

Now, we propose that GSL should be expressed as%
\begin{equation}
\dot{S}_{h}+d_{p}\dot{S}+\dot{S}_{t}\geq 0  \label{S00}
\end{equation}%
where $d_{p}\dot{S}\equiv \partial _{t}\left( d_{p}S\right) $. Summing up
Eqs. (\ref{Sh2}) and (\ref{St}), the GSL reads%
\begin{equation}
-b\dot{E}+bW_{t}\dot{V}+V\dot{\rho}_{t}+(\rho _{t}+p_{t})\dot{V}=(1-b)\dot{%
\rho}_{t}V+(1-\frac{b}{2})(\rho _{t}+p_{t})\dot{V}\geq 0.  \label{Condition0}
\end{equation}%
It should be noted that if there is no energy flow ($q_{t}=0$) between the
horizon and energy therein, the horizon and the energy inside it are in
thermal equilibrium $b=1$.

One can immediately find that the GSL holds for Einstein gravity. Two
Friedmann equations (\ref{H21}) (\ref{H22}) make the energy flow $q$ in the
continuity equation (\ref{Con2}) vanish, thereby the thermal flow vanishes.
Employing two Friedmann equations (\ref{H21}) (\ref{H22}) in condition (\ref%
{Condition0}) by replacing $\rho _{t}$ ($p_{t}$) as $\rho $ ($p$), we have
\begin{equation}
-\dot{E}+W\dot{V}+V\dot{\rho}+(\rho +p)\dot{V}=\frac{1}{2}(\rho +p)\dot{V}=%
\frac{n\left( n-1\right) }{16\pi G}\Omega _{n}H^{-n-1}\dot{H}^{2}\geq 0.
\label{EinC}
\end{equation}%
In the derivation of the above equation we have used the second Friedmann
equation $\dot{H}=-\frac{8\pi G}{(n-1)}(\rho +p)$ and $\dot{V}=\partial
_{t}(\Omega _{n}\tilde{r}_{A}^{n})=\partial _{t}(\Omega _{n}H^{-n})=-n\Omega
_{n}H^{-n-1}\dot{H}$. In \cite{Zhou}, the condition (\ref{EinC}) without the
term $W\dot{V}$ was obtained, which corresponds to the approximation $T=%
\frac{1}{2\pi \tilde{r}_{A}}$.

Now we consider a Schwarzschild black hole inside the apparent horizon in $%
\left( 3+1\right) $-dimensional spacetime, whose mass is assumed to be small
enough $M\ll\rho V$ so that the FRW metric remains unchanged (thereby $b=1$%
). Using the first Friedmann equation (\ref{H21}) in $\left( 3+1\right) $%
-dimensional spacetime, this condition reduces to%
\begin{equation}
MH\ll\frac{\tilde{r}_{A}^{3}H^{3}}{2}=\frac{1}{2},  \label{MH}
\end{equation}
where we set $G=1$. In a fluid with the energy density $\rho$ and the
pressure $p$, the change rate of the black hole mass was obtained in \cite%
{Babichev}%
\begin{equation*}
\dot{M}=4\alpha r_{h}^{2}(\rho+p)M^{2}=-4\alpha M^{2}\dot{H},
\end{equation*}
where $r_{h}$ is the radius of the black hole horizon and $\alpha\sim O(1)$
is a positive numerical constant. The entropy of the black hole is $%
S_{bl}=4\pi M^{2}$ \cite{Bekenstein1}, therefore%
\begin{equation}
\dot{S}_{bl}=-32\pi\alpha M^{2}\dot{H}.  \label{sbl}
\end{equation}
To protect the GSL, we require that the sum of the apparent horizon entropy,
the ordinary matter entropy and the black hole entropy cannot decrease with
time:%
\begin{equation*}
\dot{S}_{h}+\dot{S}_{\rho}+\dot{S}_{bl}\geq0.
\end{equation*}
Employing Eqs. (\ref{Sh2}), (\ref{Sroi}), (\ref{sbl}), together with the
Friedmann equation (\ref{H21}) and $b=1$, the condition to protect the GSL
reads%
\begin{equation*}
2\pi\dot{H}\left( -16\alpha M^{3}+\frac{\dot{H}}{2H^{5}}\frac{1}{1-\epsilon }%
\right) \geq0.
\end{equation*}
If the fluid surrounding the black hole is quintessence $\dot{H}<0$, the GSL
holds always. If the fluid is the phantom type, $\dot{H}>0$, the GSL can
also hold, because%
\begin{equation*}
\frac{\dot{H}}{H^{2}}\frac{1}{1-\epsilon}\geq32\alpha M^{3}H^{3}\sim0,
\end{equation*}
where the condition (\ref{MH}) and the positive temperature has been
considered.

In the following, we are going to extend the discussion on the condition (%
\ref{Condition0}) to protect the GSL to generalized gravity theories.

\subsection{Lovelock gravity}

The Lagrangian of the Lovelock gravity consists of the dimensionally
extended Euler densities \cite{Lovelock}%
\begin{equation*}
L=\sum_{i=1}^{[n/2]}c_{i}L_{i},
\end{equation*}%
where $c_{i}$ is an arbitrary positive constant and $L_{i}$ is the Euler
density of a ($2i$)-dimensional manifold%
\begin{equation*}
L_{i}=2^{-i}\delta _{\alpha _{1}\beta _{1}\cdots \alpha _{ii}\beta
_{i}}^{\mu _{1}\nu _{1}\cdots \mu _{i}\nu _{i}}R_{\mu _{1}\nu _{1}\cdots \mu
_{i}\nu _{i}}^{\alpha _{1}\beta _{1}\cdots \alpha _{ii}\beta _{i}}.
\end{equation*}%
$L_{1}$ is just the Einstein-Hilbert term, and $L_{2}$ corresponds to the so
called Gauss-Bonnet term. Using the FRW metric, we obtain Friedmann
equations in ($n+1$)-dimensional spacetime%
\begin{equation}
\sum_{i=1}^{[n/2]}\hat{c}_{i}\left( H^{2}\right) ^{i}=\frac{16\pi G}{n(n-1)}%
\rho ,  \label{FM1Lovelock}
\end{equation}%
and%
\begin{equation}
\sum_{i=1}^{[n/2]}\hat{c}_{i}i\left( H^{2}\right) ^{i-1}(\dot{H})=-\frac{%
8\pi G}{(n-1)}(\rho +p),  \label{LoveH2}
\end{equation}%
where%
\begin{equation*}
\hat{c}_{i}=\frac{(n-2)!}{(n-2i)!}c_{i}
\end{equation*}%
Since only one dynamic field $\rho $ in the first Friedmann equation (\ref%
{FM1Lovelock}), the first law on the horizon is described by the equilibrium
thermodynamics $d_{p}S=0$. Two Friedmann equations (\ref{FM1Lovelock}) (\ref%
{LoveH2}) make the energy flow $q=0$, so the horizon and energy are in
thermodynamical equilibrium $b=1$. From two Friedmann equations (\ref%
{FM1Lovelock}) (\ref{LoveH2}), one can find that the condition (\ref%
{Condition0}) always holds%
\begin{equation}
\frac{1}{2}(\rho +p)\dot{V}=\frac{n\left( n-1\right) }{16\pi G}\Omega
_{n}H^{-1-n}\sum_{i=1}^{[n/2]}\hat{c}_{i}i\left( H^{2}\right) ^{i-1}\dot{H}%
^{2}\geq 0.  \label{GSllovelock}
\end{equation}%
The condition to protect the GSL in the Einstein gravity (\ref{EinC}) can be
recovered when $\hat{c}_{1}=1$, $\hat{c}_{i}=0$ ($i>1$).

\subsection{Randall-Sundrum braneworld gravity}

We consider a $n$-dimensional brane embedded in a ($n+2$)-dimensional
spacetime. Using the junction condition on the brane, we can obtain
Friedmann equations%
\begin{equation}
H^{2}=\frac{1}{4n^{2}}\rho^{2}+\frac{1}{2n^{2}}\lambda\rho
\label{RSFriedmann1}
\end{equation}%
\begin{equation}
\dot{H}=-\frac{1}{4n}(\rho+p)(\rho+\lambda)  \label{RSFriedmann2}
\end{equation}
where the Randall-Sundrum fine-turning condition%
\begin{equation*}
\frac{1}{4n^{2}}\lambda^{2}+\frac{2\Lambda_{n+2}}{n(n+1)}=0
\end{equation*}
has been used. Using the junction condition into the $(05)$ component of the
field equation we can obtain the conserved equation%
\begin{equation}
\dot{\rho}+nH(\rho+p)=q=0,  \label{RSq3}
\end{equation}
which results that the horizon and energy inside the horizon are in
thermodynamical equilibrium $b=1$.

We see that $\rho $ is the only freedom in the first Friedmann equation (\ref%
{RSFriedmann1}), so we only need to consider the equilibrium thermodynamics
about the horizon. From (\ref{RSFriedmann1}) and (\ref{RSFriedmann2}), one
can find that the condition (\ref{Condition0}) to protect the GSL always
holds%
\begin{equation}
\frac{1}{2}(\rho +p)\dot{V}=\frac{2n^{2}\Omega _{n}}{\sqrt{\lambda
^{2}+4n^{2}H^{2}}}H^{-n-1}\dot{H}^{2}\geq 0,  \label{GSLbrane}
\end{equation}%
The condition (\ref{GSLbrane}) can be reduced to the condition in Einstein
gravity (\ref{EinC}) at low energy $\rho \ll \lambda $.

\subsection{Nonlinear gravity}

\bigskip For the nonlinear gravity $f(R)$, the Lagrangian is%
\begin{equation*}
L=\frac{1}{16\pi G}f(R)
\end{equation*}%
The variational principle gives equations of motion. Using the FRW metric,
one can obtain Friedmann equations in ($n+1$)-dimensional space-time%
\begin{equation}
H^{2}=\frac{16\pi G}{n(n-1)}\frac{1}{f^{\prime }}\left( \rho +\rho
_{c}f^{\prime }\right)  \label{FMfr1}
\end{equation}%
\begin{equation}
\dot{H}=-\frac{8\pi G}{(n-1)}\frac{1}{f^{\prime }}(\rho +\rho _{c}f^{\prime
}+p+p_{c}f^{\prime }),  \label{FMfr2}
\end{equation}%
where%
\begin{equation*}
\rho _{c}=\frac{1}{8\pi Gf^{\prime }}\left[ -\frac{f-Rf^{\prime }}{2}%
-nHf^{\prime \prime }\dot{R}\right]
\end{equation*}%
\begin{equation*}
p_{c}=\frac{1}{8\pi Gf^{\prime }}\left[ (f-Rf^{\prime })-f^{\prime \prime }%
\ddot{R}+f^{\prime \prime \prime }\dot{R}^{2}+n(n-1)f^{\prime \prime }\dot{R}%
\right] .
\end{equation*}%
The prime denotes the derivate respect to $R$. Since $\frac{\partial H^{2}}{%
\partial \rho }$ is determined by dynamic field $f^{\prime }$ while not the
horizon radius uniquely, one should consider the horizon described by
non-equilibrium thermodynamics. One can select $\rho _{p}$ arbitrarily. For
example, we select $\rho _{p}=(f^{\prime },\rho _{c})$. There is not the
real matter field besides the ordinary matter $\rho $. It is important to
observe that from the Friedmann equations, $\rho _{e}\equiv \rho
_{c}f^{\prime }$ ($p_{e}\equiv p_{c}f^{\prime }$) acts as the density
(pressure) of an effective energy component in $f(R)$ gravity.

The condition (\ref{Condition0}) reads%
\begin{equation*}
(1-b)\dot{\rho}_{t}V+(1-\frac{b}{2})(\rho _{t}+p_{t})\dot{V}\geq 0,
\end{equation*}%
where the total density (pressure) is $\rho _{t}\equiv \rho +\rho _{e}$ ($%
p_{t}\equiv p+p_{e}$). Solving $\rho _{t}$ and $p_{t}$ from two Friedmann
equations (\ref{FMfr1}) and (\ref{FMfr2}), and substituting them into the
above inequality, we find%
\begin{equation}
\frac{n(n-1)\Omega _{n}}{16\pi G}H^{-(n+1)}\left[ (1-b)H^{3}\dot{f}^{\prime
}+2(1-b)f^{\prime }H^{2}\dot{H}+(2-b)f^{\prime }\dot{H}^{2}\right] \geq 0.
\label{GSLfr0}
\end{equation}%
This inequality is the condition to protect the GSL in the $f(R)$ gravity.
Now we like to point out a nontrivial observation that this condition may
hold always. From two Friedmann equations (\ref{FMfr1}) and (\ref{FMfr2}),
we find that $\dot{f}^{\prime }$ is related to the total energy flow $q_{t}$%
\begin{equation}
q_{t}=\frac{n(n-1)}{16\pi G}H^{2}\dot{f}^{\prime }.  \label{qtt}
\end{equation}%
Since $G/f^{\prime }$ takes role as the effective Newton gravitational
constant, the relation (\ref{qtt}) means the energy fluid is determined by
the evolvement of effective Newton gravitational constant. Consider the
temperature of total energy $bT$, which in general can not equal to the
temperature of horizon $T$. However, it is known that the experimental
bounds acquires the Newton constant should be approximate constant \cite%
{Ozan}. From the relation (\ref{qtt}), one can find that the big energy
fluid is prohibited to protect the effective Newton constant as approximate
constant. Thereby the thermodynamic fluid is small and the temperature of
the horizon is very closed with the one of the energy source therein $b\sim 1
$. This is more reasonable if two systems undergo some length of time. So
the former two terms in the bracket of condition (\ref{GSLfr0}) may be
neglected. Moreover, since the gravity is always attractive, we can impose $%
f^{\prime }>0$. Thus the last term in the bracket is positive and the GSL
always holds. We also note that when the effective Newton constant is
constant indeed $\dot{f}^{\prime }=0$ which leads $q_{t}=0$ then $b=1$, the
GSL always holds, recovering the results of Einstein gravity.

\bigskip It is interesting to show the entropy production $d_{i}S$ clearly.
Recalling the conserved equation (\ref{Con1})%
\begin{equation}
\dot{\rho}_{eff}=-nH(\rho _{eff}+p_{eff})  \label{con}
\end{equation}%
and using the first Friedmann equation (\ref{FMfr1}), one can find that the
l.h.s in Eq. (\ref{con}) reads%
\begin{align}
\dot{\rho}_{eff}& =\frac{n(n-1)}{16\pi G}\left( \frac{\partial H^{2}}{%
\partial \rho }\dot{\rho}+\frac{\partial H^{2}}{\partial \rho _{p}}\dot{\rho}%
_{p}\right)   \notag \\
& =\frac{n(n-1)}{16\pi G}\left( \frac{16\pi G}{n(n-1)}\frac{1}{f^{\prime }}%
\dot{\rho}+\frac{\partial H^{2}}{\partial \rho _{p}}\dot{\rho}_{p}\right)
\notag \\
& =\frac{1}{f^{\prime }}\dot{\rho}+\frac{n(n-1)}{16\pi G}\frac{\partial H^{2}%
}{\partial \rho _{p}}\dot{\rho}_{p}  \label{r1}
\end{align}%
while by using the second Friedmann equation (\ref{FMfr2}) the r.h.s in Eq. (%
\ref{con}) reads
\begin{equation}
-nH(\rho _{eff}+p_{eff})=-nH\left[ \frac{1}{f^{\prime }}\left( \rho
_{t}+p_{t}\right) \right] .  \label{r2}
\end{equation}%
Employing the continuous equation (\ref{Con2}) to Eqs. (\ref{r1}) and (\ref%
{r2}), one can find%
\begin{equation}
\frac{\partial H^{2}}{\partial \rho _{p}}\dot{\rho}_{p}=-\frac{16\pi G}{%
n(n-1)}\frac{1}{f^{\prime }}\left[ nH(\rho _{t}+p_{t}-\rho -p)+q\right] .
\label{ls}
\end{equation}%
Using the continuous equation (\ref{qt}) and substituting Eq. (\ref{ls})
into Eq. (\ref{dE}), we have%
\begin{eqnarray}
\delta Q &=&nVH(\rho +p)dt-Vqdt-Vdt\frac{1}{\frac{\partial H^{2}}{\partial
\rho }}\frac{\partial H^{2}}{\partial \rho _{p}}\dot{\rho}_{p}-V\frac{1}{%
\frac{\partial H^{2}}{\partial \rho }}\frac{\dot{H}^{2}}{H}+T\frac{16\pi G}{%
n(n-1)}S_{E}d\frac{1}{\frac{\partial H^{2}}{\partial \rho }}  \notag \\
&=&nVH(\rho _{t}+p_{t})dt-Vdt\frac{1}{\frac{\partial H^{2}}{\partial \rho }}%
\frac{\dot{H}^{2}}{H}+TS_{E}df^{\prime }  \notag \\
&=&nVH(\rho _{t}+p_{t})dt-\frac{1}{2}\left( \rho _{t}+p_{t}\right) \dot{V}%
+TS_{E}df^{\prime }  \notag \\
&=&-dE+W_{t}dV+Vq_{t}dt+TS_{E}df^{\prime }.  \label{dqq}
\end{eqnarray}%
One can find clearly that the energy fluid $Vq_{t}dt$ accounts for some part
of the thermodynamical fluid. From Eqs. (\ref{dis}) and (\ref{dqq}), the
entropy production can be obtained
\begin{eqnarray*}
d_{i}S &=&-\frac{1}{T}Vq_{t}dt-S_{E}df^{\prime } \\
&=&\frac{-n\Omega _{n}H^{-n+1}df^{\prime }\left[ \left( n+1\right) H^{2}+%
\dot{H}\right] }{4G(2H^{2}+\dot{H})}.
\end{eqnarray*}%
The second equation has used the two Friedmann equations and Eq. (\ref{qtt}%
). One can find $d_{i}S$ is not an exact form as desired and it is vanishing
when the effective Newton gravitational constant $f^{\prime }$ is constant
indeed. This is a reasonable result since the Einstein gravity is recovered
in this situation. It should be noticed that this entropy production is
different with the one given in Ref. \cite{Eling,Cao}, where they have not
considered the energy flux of effective energy $\rho _{e}$. If we really
omit the energy flux of effective energy, the condition to protect GSL (\ref%
{S00}) is (for simplicity, we consider the ($3+1$)-dimensional space-time)%
\begin{eqnarray*}
&&\dot{S}_{h}+d_{i}\dot{S}+\dot{S}_{t}=-b\left[ \partial _{t}(\rho V)+W\dot{V%
}\right] +V\dot{\rho}_{t}+(\rho _{t}+p_{t})\dot{V} \\
&=&\frac{\Omega _{3}}{32\pi GH^{4}}\left\{
\begin{array}{c}
12(1+b)H^{3}\dot{f}^{\prime }-3bf\dot{H}+6(4+b)f^{\prime }\dot{H}^{2} \\
+12H^{2}\left[ (2+5b)f^{\prime }\dot{H}-b\ddot{f}^{\prime }\right]
-2bH\left( \dot{f}+9\dot{f}^{\prime }\dot{H}-6f^{\prime }\ddot{H}\right)
\end{array}%
\right\} \geq 0,
\end{eqnarray*}%
Obviously, the GSL can not be held always in this situation, which is a
strong point to favor the present entropy production (\ref{dis}).

\subsection{Scalar-tensor gravity}

The general scalar-tensor theory of gravity is described by the Lagrangian%
\begin{equation*}
L=F\left( \phi \right) R-\frac{1}{2}g^{\mu \nu }\partial _{\mu }\phi
\partial _{\nu }\phi -V\left( \phi \right) ,
\end{equation*}%
where $F(\phi )$ is a positive continuous function of the scalar field $\phi
$ and $V(\phi )$ is its potential. Using the FRW metric, we obtain Friedmann
equations in ($n+1$)-dimensional space-time%
\begin{equation}
H^{2}=\frac{16\pi G}{n(n-1)}\frac{1}{F}\left( \rho +\rho _{f}+\rho
_{c}F\right)   \label{FMfai1}
\end{equation}%
\begin{equation}
\dot{H}=\frac{8\pi G}{(n-1)}\frac{1}{F}\left( \rho +p+\rho _{f}+p_{f}+\rho
_{c}F+p_{c}F\right) .  \label{FMfai2}
\end{equation}%
where the density and pressure of scalar field $\phi $ are%
\begin{align*}
\rho _{f}& =\frac{1}{2}\dot{\phi}^{2}+V\left( \phi \right)  \\
p_{f}& =\frac{1}{2}\dot{\phi}^{2}-V\left( \phi \right) ,
\end{align*}%
and $\rho _{e}\equiv \rho _{c}F$ ($p_{e}\equiv p_{c}F$) can be understood as
effective density (pressure) of the energy component in scalar-tensor
theory:
\begin{equation*}
\rho _{c}=-\frac{n}{8\pi GF}H\dot{F}
\end{equation*}%
\begin{equation*}
p_{c}=\frac{1}{8\pi GF}\left[ \ddot{F}+\left( n-1\right) H\dot{F}\right] .
\end{equation*}%
Obviously, we need to consider the non-equilibrium thermodynamics of
horizon. We select $\rho _{p}=(\rho _{f},F,\rho _{e})$.

Consider the condition (\ref{Condition0}) for GSL%
\begin{equation*}
(1-b)\dot{\rho}_{t}V+(1-\frac{b}{2})(\rho _{t}+p_{t})\dot{V}\geq 0,
\end{equation*}%
where the total density (pressure) is $\rho _{t}\equiv \rho +\rho _{f}+\rho
_{e}$ ($p_{t}\equiv p+p_{f}+p_{e}$). Solving $\rho _{t}$ and $p_{t}$ from
two Friedmann equations (\ref{FMfai1}) and (\ref{FMfai2}), and substituting
them into the above inequality, we find that the inequality can be reduced
to,%
\begin{equation}
\frac{n(n-1)\Omega _{n}}{16\pi G}H^{-(n+1)}\left[ (1-b)H^{3}\dot{F}%
+2(1-b)FH^{2}\dot{H}+(2-b)F\dot{H}^{2}\right] \geq 0  \label{GSLScalar0}
\end{equation}%
The effective Newton gravitational constant in the scalar tensor theory is
taken as $G/F$. From two Friedmann equations (\ref{FMfai1}) and (\ref{FMfai2}%
), the dynamical Newton constant is related to the energy flow%
\begin{equation}
q_{t}=\frac{n(n-1)}{16\pi G}H^{2}\dot{F}.  \label{qt2}
\end{equation}%
The condition (\ref{GSLScalar0}) and the relation (\ref{qt2}) are the same
as those in the $f(R)$ gravity. As observed in the $f(R)$ gravity, we can
think that the GSL for scalar-tensor theory always holds, since we can
impose $F>0$ which means the gravity is always attractive, and the big
energy fluid is prohibited to protect the effective Newton constant as
approximate constant under experimental bounds, thereby the temperature of
the horizon is very closed with the one of the energy source therein $b\sim
1 $. Moreover, it should be noticed that, actually $f(R)$ gravity is a
special scalar-tensor theory by introducing the scalar field $\phi =R$ and
potential $V=\phi f^{\prime }-f$ and choosing the Brans-Dick parameter $%
\omega =0$ (\cite{Brans,Bergmann}---see \cite{Faraoni} for a review).

Now we will evaluate the entropy production. The process is similar to the
one of nonlinear gravity. Recalling the continuous equation (\ref{Con1}) and
using the first Friedmann equation (\ref{FMfai1}), one can find that the
l.h.s in Eq. (\ref{con}) reads
\begin{equation}
\dot{\rho}_{eff}=\frac{1}{F}\dot{\rho}+\frac{n(n-1)}{16\pi G}\frac{\partial
H^{2}}{\partial \rho _{p}}\dot{\rho}_{p}.  \label{r3}
\end{equation}%
Using the second Friedmann equation (\ref{FMfai2}) the r.h.s in Eq. (\ref%
{con}) reads
\begin{equation}
-nH(\rho _{eff}+p_{eff})=-nH\frac{1}{F}(\rho _{t}+p_{t}).  \label{r4}
\end{equation}%
Employing the continuous equation (\ref{Con2}) to Eqs. (\ref{r3}) and (\ref%
{r4}), one can find%
\begin{equation}
\frac{\partial H^{2}}{\partial \rho _{p}}\dot{\rho}_{p}=-\frac{16\pi G}{%
n(n-1)}\frac{1}{F}\left[ nH(\rho _{t}+p_{t}-\rho -p)+q\right] .  \label{ls1}
\end{equation}%
By substituting Eq. (\ref{ls1}) into Eq. (\ref{dE}), one can obtain
\begin{equation*}
\delta Q=-dE+W_{t}dV+Vq_{t}dt+TS_{E}dF.
\end{equation*}%
Thus the entropy production can be obtained from Eqs. (\ref{dis}) and (\ref%
{qt2})
\begin{equation*}
d_{i}S=\frac{-n\Omega _{n}H^{-n+1}dF\left[ \left( n+1\right) H^{2}+\dot{H}%
\right] }{4G(2H^{2}+\dot{H})}.
\end{equation*}%
One can find $d_{i}S$ is not an exact form and it is vanishing when the
effective Newton gravitational constant $F$ is constant indeed.

\section{Summary}

In this paper, we have examined the GSL in generalized theories of gravity.
We have adopted the procedure developed in \cite{Wu1} to obtain the entropy
on the horizon. In studying the GSL, we have examined the evolution of the
entropy contributed by all matter, field and energy contents. We have
derived a universal condition to protect the GSL and examined its validity.
In Einstein gravity, (even in the phantom-dominated universe with a
Schwarzschild black hole), Lovelock gravity, and braneworld gravity, we show
that the condition to keep the GSL can always be satisfied. In $f(R)$
gravity and scalar-tensor gravity, the condition to protect the GSL can also
hold under the consideration that the gravity is always attractive and the
energy fluid between the horizon and total energy source therein is very
small to protect the effective Newton constant as the approximate constant
satisfying the experimental bounds. The same requirement for the $f(R)$
gravity to hold the GSL as that of the scalar-tensor gravity shows again
that $f(R)$ gravity is a special scalar-tensor theory.

\section*{Acknowledgment}

This work was partially supported by the NSFC, Shanghai Education
Commission, Science and Technology Commission. This work was also supported
by the NSFC under Grant Nos. 10575068 and 10604024, the Shanghai Research
Foundation No. 07dz22020, the CAS Knowledge Innovation Project Nos.
KJcx.syw.N2, the Shanghai Education Development Foundation, and the
Innovation Foundation of Shanghai University.

\end{document}